%% file: paper.tex
\pdfoutput=1
\documentclass[10pt,conference,letterpaper]{IEEEtran}

\usepackage[noadjust]{cite}
\usepackage[T1]{fontenc}
\usepackage[utf8]{inputenc}
\usepackage{balance}
\usepackage{url}
\usepackage{subfig}
\usepackage{tabularx}
\usepackage{hhline}
\usepackage{colortbl}  
\usepackage{dblfloatfix}  

\newcommand{\docauthor}{Jonathan Will\IEEEauthorrefmark{1}, Lauritz Thamsen\IEEEauthorrefmark{2}, Jonathan Bader\IEEEauthorrefmark{1}, Dominik Scheinert\IEEEauthorrefmark{1}, and Odej Kao\IEEEauthorrefmark{1}}
\newcommand{\docsubject}{\IEEEauthorrefmark{1}Technische Universit\"at Berlin, Germany \hspace{6mm} \IEEEauthorrefmark{2}University of Glasgow, United Kingdom}
\newcommand{\dockeywords}{Scalable Data Analytics, Distributed Dataflows, Profiling, Resource Allocation, Cluster Management}
\newcommand{\doctitle}{Get Your Memory Right:\\ The Crispy Resource Allocation Assistant\\for Large-Scale Data Processing}

\PassOptionsToPackage{bookmarks=false}{hyperref}
\usepackage[pdftex]{hyperref}
\hypersetup{
    pdfborder={0 0 0},
    pdfauthor=\docauthor,
    pdftitle=\doctitle,
    pdfsubject=\docsubject,
    pdfkeywords=\dockeywords
}

\usepackage{amsmath,amssymb,amsfonts}
\usepackage{algorithmic}
\usepackage{graphicx}
\usepackage{textcomp}
\usepackage{tikz}
\usepackage{xcolor}  

\definecolor{newgreen}{RGB}{7, 160, 25}
\newenvironment{gn}{\color{newgreen}}

\definecolor{Gray}{gray}{0.9}

\def\mycopyrightnotice{
  {\footnotesize 978-1-6654-9115-0/22/\$31.00~\copyright~2022 IEEE\\
   DOI: \href{https://doi.org/10.1109/IC2E55432.2022.00014}{https://doi.org/10.1109/IC2E55432.2022.00014}
  }
  \gdef\mycopyrightnotice{}
}
\makeatletter
  \def\ps@IEEEtitlepagestyle{%
 \def\@oddfoot{\mycopyrightnotice}
  \def\@evenfoot{}
  }%
\makeatother

\def\BibTeX{{\rm B\kern-.05em{\sc i\kern-.025em b}\kern-.08em T\kern-.1667em\lower.7ex\hbox{E}\kern-.125emX}}

\newcommand*\circled[1]{\tikz[baseline=(char.base)]{
            \node[shape=circle,draw,inner sep=1.5pt] (char) {#1};}}

\begin{document}

\title{\doctitle}

\author{%
\IEEEauthorblockN{\docauthor}
\IEEEauthorblockA{\docsubject\\
\{will, jonathan.bader, dominik.scheinert, odej.kao\}@tu-berlin.de \hspace{6mm} lauritz.thamsen@glasgow.ac.uk
}}

\maketitle

\begin{abstract}
\input{sections/abstract.tex}
\end{abstract}

\IEEEpeerreviewmaketitle


\section{Introduction}\label{sec:INTRO}
\input{sections/1_introduction}

\section{Problem Analysis}\label{sec:PROBLEM_ANALYSIS}
\input{sections/3_problem_analysis}

\section{Approach}\label{sec:APPROACH}
\input{sections/4_approach}

\section{Evaluation}\label{sec:EVALUATION}
\input{sections/5_evaluation}

\section{Related Work}\label{sec:RELATED_WORK}
\input{sections/6_related_work}

\section{Conclusion}\label{sec:CONCLUSION}
\input{sections/7_conclusion}

\section*{Acknowledgments}

This work has been supported through grants by the German Ministry for Education and Research (BMBF) as BIFOLD (grant 01IS18025A) and the German Research Foundation (DFG) as FONDA (DFG Collaborative Research Center 1404).

\bibliographystyle{IEEEtran}
\balance
\bibliography{./references}

\end{document}

%% file: sections/abstract.tex
Distributed dataflow systems like Apache Spark and Apache Hadoop enable data-parallel processing of large datasets on clusters.
Yet, selecting appropriate computational resources for dataflow jobs –– that neither lead to bottlenecks nor to low resource utilization –– is often challenging, even for expert users such as data engineers.
Further, existing automated approaches to resource selection rely on the assumption that a job is recurring to learn from previous runs or to warrant the cost of full test runs to learn from.
However, this assumption often does not hold since many jobs are too unique.

Therefore, we present \textit{Crispy}, a method for optimizing data processing cluster configurations based on job profiling runs with small samples of the dataset on just a single machine.
Crispy attempts to extrapolate the memory usage for the full dataset to then choose a cluster configuration with enough total memory.
In our evaluation on a dataset with 1031 Spark and Hadoop jobs, we see a reduction of job execution costs by 56\% compared to the baseline, while on average spending less than ten minutes on profiling runs per job on a consumer-grade laptop.

%% file: sections/1_introduction.tex
Many organizations today have to analyze large amounts of data.
This applies to companies, public sector organizations, and scientific computing.
Example application areas include marketing, public infrastructure monitoring, bioinformatics, and geosciences~\cite{hafez2016effective,geldenhuys2021dependable,bader2021tarema,bader2022lotaru,lehmann2021force}.
Distributed dataflow systems like Apache Spark~\cite{spark} and Apache Flink~\cite{flink} simplify developing scalable data-parallel programs, reducing especially the need to implement parallelism and fault tolerance.

However, it is often not straightforward to select resources and configure clusters for efficiently executing such programs~\cite{lama2012aroma, rajan2016perforator}.
This is especially the case for users who only infrequently run large-scale data processing jobs and cannot rely on abundant help from systems operations staff.
To avoid resource bottlenecks and ensure performance expectations are met, users typically overprovision resources, leading to unnecessarily high costs and low resource utilization~\cite{yang2013bubble,liu2011measurement,delimitrou2014quasar,lin2013scaling}.
These issues are amplified when using larger cluster setups.
For instance, suboptimal resource configurations can increase costs tenfold when using public clouds~\cite{cherrypick, hsu2018arrow}.

Several approaches for automated cluster resource selection build runtime models from historical executions or from dedicated profiling runs to evaluate possible configurations~\cite{ernest, hongzi2016resource, chen2021silhouette, baughman2018profiling, rajan2016perforator, bell, scheinert2021enel, scheinert2021bellamy, scheinert2021potential, will2020towards, will2021c3o, will2021training, yadwadkar2017selecting}.
Several other approaches iteratively search for suitable cluster configurations~\cite{cherrypick, hsu2018micky, hsu2018arrow, fekry2020accelerating, he2019statistics}.
That is, these approaches involve significant overhead for testing configurations, which is offset by expected resource efficiency gains in future executions.
The assumption of many jobs being recurring is supported by reports from organizations regarding their cluster usage.
For instance, inquiries by Microsoft and Alibaba found that 40\%-65\% of data analytics jobs on their clusters were recurring~\cite{agarwahl2012reoptimizing, jyothi2016morpheus, wang2020grosbeak}.
Conversely, however, a significant amount of jobs are unique, one-off executions without significant similarity to previously executed jobs.
The aforementioned approaches for configuring a resource-efficient cluster lose their efficacy in those scenarios due to their expensive data gathering and search costs.

In this paper, we present \emph{Crispy}, a method for efficiently optimizing data processing cluster configurations, which does not assume that jobs run repeatedly.
This approach is based on our analysis of a dataset with 1031 Spark and Hadoop jobs that were executed on Amazon Web Services (AWS)~\cite{hsu2018arrow}.
Our results indicate that a particularly crucial part of selecting computational resources for efficient dataflow jobs is allocating enough total memory to prevent expensive and avoidable repeated read operations from storage or recomputations.
The resource selections of our approach are thus based on estimations of cluster memory needs and, therefore, the avoidance of both expensive memory bottlenecks and overprovisioning.

Crispy conducts multiple job runs on small samples of the dataset and extrapolates the memory usage for the full dataset to choose a cluster with just enough total memory.
These sample runs can be executed on a single machine, for instance, a developer's laptop or desktop PC\@, and they are completed within minutes.
If Crispy finds a linear relation between input data size and memory usage, it assumes that the memory requirement can be successfully estimated for the full dataset.
Otherwise, Crispy falls back to a robust baseline approach.
\vspace{2mm}
\\
\emph{Contributions}. The contributions\footnote{\href{https://github.com/dos-group/crispy}{github.com/dos-group/crispy}} of this paper are:
\vspace{0mm}
\begin{itemize}
    \item An analysis of the impact of total cluster memory allocation on resource efficiency for distributed dataflow jobs
    \item A method for quickly and resource-efficiently profiling data processing jobs to model memory requirements
    \item A method for selecting cluster configurations for distributed data processing jobs according to memory needs
\end{itemize}
\vspace{-3mm}

\begin{figure*}[bh]
\vspace{-4mm}
    \subfloat{%
      \includegraphics[width=.33\linewidth]{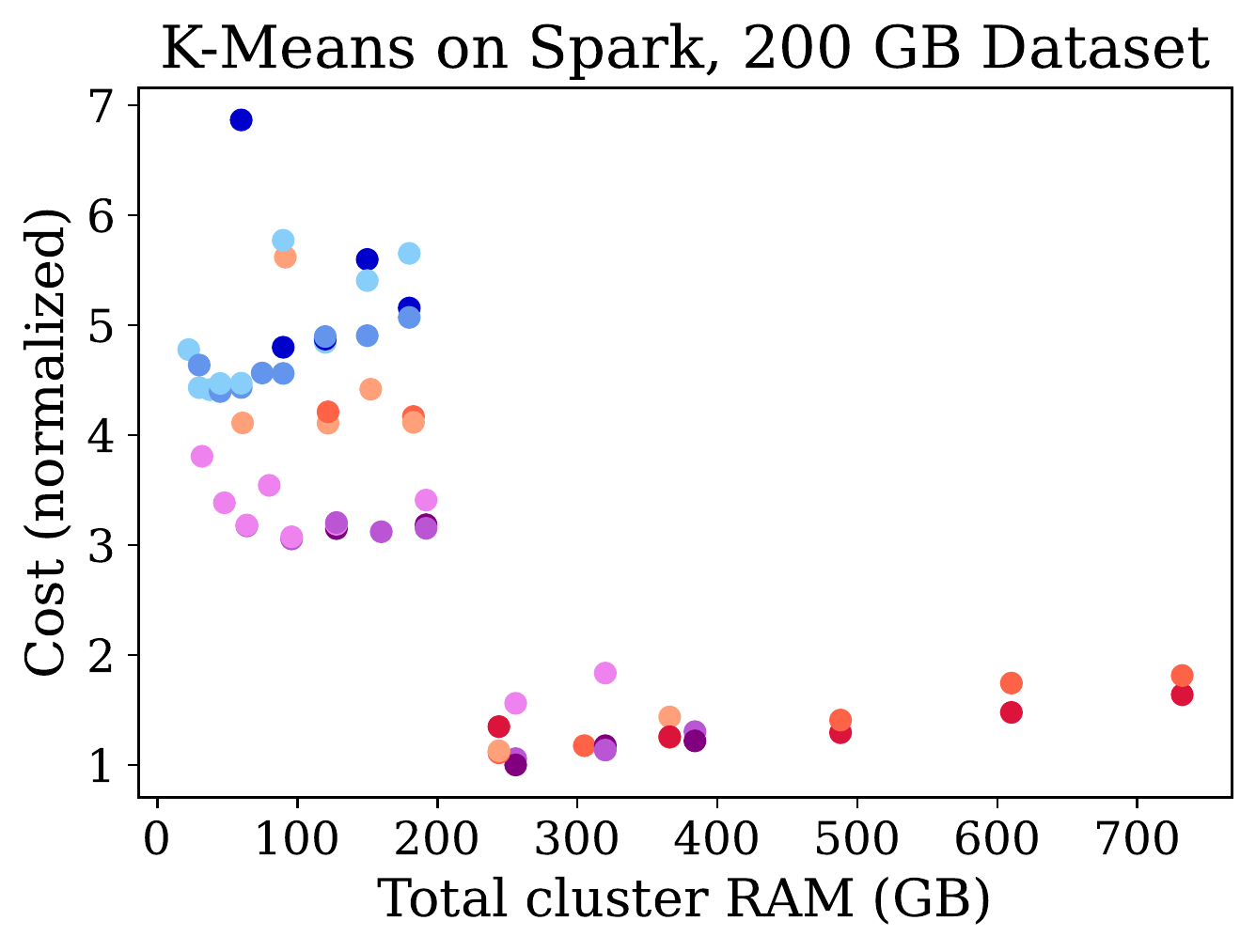}%
    }\hfill
    \subfloat{%
      \includegraphics[width=.33\linewidth]{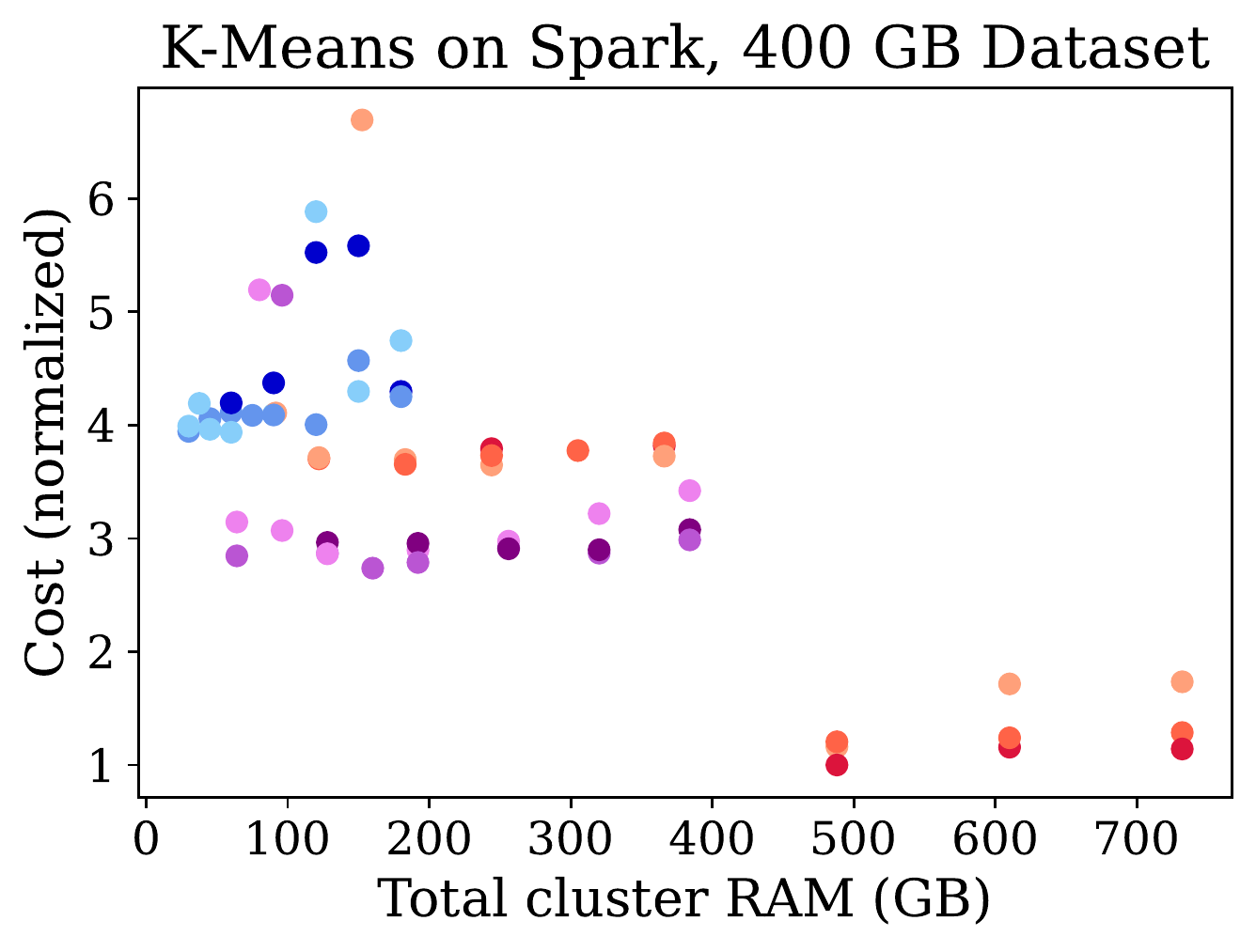}%
    }\hfill
    \subfloat{%
      \includegraphics[width=.34\linewidth]{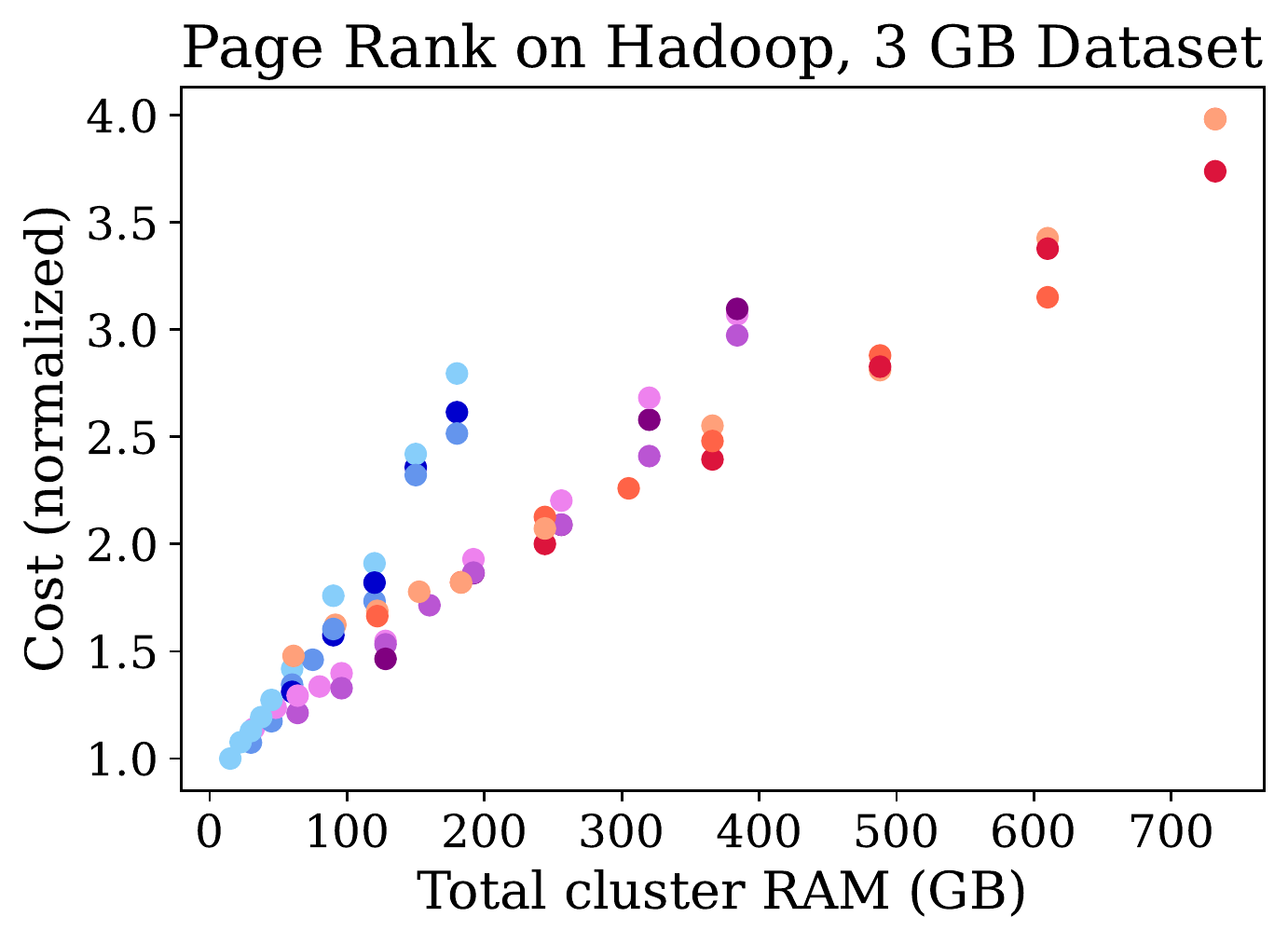}%
    }\hfill
    \centering
    \vspace{-4.0mm}
    \subfloat{\includegraphics[width=.4\textwidth]{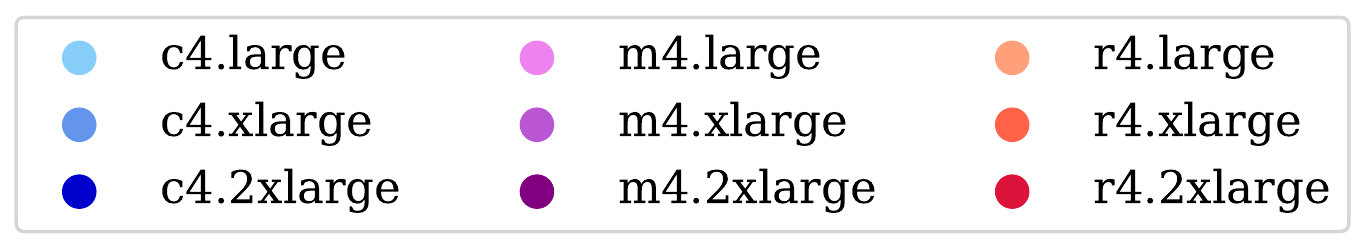}}
    \vspace{-1.3mm}
    \caption{Total cluster RAM versus monetary cost for selected jobs with different AWS machine types at different scale-outs.}\label{fig:memory_bottlenecks}
\end{figure*}

\iftrue
\emph{Outline}. The remainder of the paper is structured as follows.
Section~\ref{sec:PROBLEM_ANALYSIS} analyses the problem of selecting cluster resources for data processing.
Section~\ref{sec:APPROACH} presents the approach of the Crispy cluster resource allocator.
Section~\ref{sec:EVALUATION} evaluates Crispy.
Section~\ref{sec:RELATED_WORK} discusses related work.
Section~\ref{sec:CONCLUSION} summarizes and concludes this paper.
\fi

%% file: sections/3_problem_analysis.tex
This section introduces the problem of efficient resource allocation for distributed dataflow systems while focusing on memory sufficiency.

\subsection{Cluster Resource Selection for Distributed Dataflow Jobs}

Distributed dataflows are graphs of connected data-parallel operators that execute user-defined functions on a set of shared-nothing commodity cluster nodes.
Users can easily create data-parallel programs through high-level programming abstractions without needing to handle the parallelization explicitly.
The system translates a user's sequential program into a directed graph of data-parallel operators and finally into an optimized execution plan.
Such systems also handle failures by automatically repeating failed operations and replacing defective nodes.

A significant improvement of newer distributed dataflow frameworks like Apache Spark~\cite{spark} and Apache Flink~\cite{flink} over the older Hadoop MapReduce~\cite{mapreduce} is the caching of data in memory for faster read access.
The user or the framework itself can choose strategies for data that could not fit into memory, e.g., spilling it to disk or recomputing the data from previous stages on demand.
One supporting technique is to serialize data upon caching to save memory space at the expense of some overhead from (de)serialization.

A data processing cluster consists of several nodes working in parallel.
Often, the nodes in a cluster are virtual machines, which vary principally in their number of cores and the amount of memory per core, but some can also offer additional I/O speed or network capabilities.
In AWS, for instance, virtual machines of the \emph{c} type have less memory per core than those of the \emph{r} type, while machines of the \emph{m} type lie between those two.
Denominations like \emph{large}, \emph{xlarge} and \emph{2xlarge} refer to the amount of cores per machine.
Users of shared private clusters have to make similar considerations when allocating resources for individual jobs.

Regarding the amount of individual allocated resources, there is generally a trade-off between speed and cost of execution.
The different costs of individual resources like CPU cores and memory are reflected in the prices for virtual machines in public clouds, while in a private shared cluster, some resource categories can be abundant and others scarce, depending on current usage.
Hence, to increase performance efficiently, the challenge is to add the resources with the most cost-efficient performance yield.

\subsection{Memory Bottlenecks in Distributed Dataflow Jobs}

For many iterative jobs, e.g., machine learning jobs like K-Means and Page Rank, the whole dataset is read at every iteration.
To avoid repeated disk read operations or recomputations, the full dataset needs to be retained in the combined cluster memory.
If a lack of memory prevents this, the job exhibits a significant slowdown, a \emph{memory bottleneck}.

In Figure~\ref{fig:memory_bottlenecks}, we see an example of memory bottlenecks for K-Means jobs on Spark.
Here, a marginal increase in total cluster memory can lead to the dataset fitting into memory,
causing drastically reduced runtime and thereby leading to a lower job execution cost.
In contrast, memory sufficiency is no issue for Page Rank on Hadoop due to the lack of caching. The data is read from disk at every iteration in any case.

Having enough memory for caching is crucial for preventing overhead, while additional memory is ineffective in speeding up the execution. 
In contrast, the performance increase for additional resources like CPU cores is more gradual.
Consequently, an automated resource allocation assistant must help the user avoid memory bottlenecks and thereby avoid costly but slow executions.
Once possible memory bottlenecks are mitigated, adding more or fewer resources of other types, especially CPU cores, becomes a simpler cost-performance trade-off that can be up to the user to decide, determined by current resource costs and user preference.

%% file: sections/4_approach.tex
This section presents our approach to the problem of finding the best cluster configuration for a distributed dataflow job.
We first present the overall idea of the method.
Then, we explain how job profiling and memory usage estimation can help us select good cluster resources.

\subsection{Overview}

The goal is to select suitable cluster resources for data processing jobs without assuming recurrence.
This imposes the limitation that we cannot use execution metrics from past job executions.
Likewise, we cannot let the full job run on different clusters and have this search cost be amortized by more resource-efficient future job executions.

Therefore, we attempt to gather enough information about the resource requirements of a job by executing it on a small sample of the dataset on a single machine, e.g., the developer's laptop.
In accordance with our results from the problem analysis in Section~\ref{sec:PROBLEM_ANALYSIS}, we are most interested in the job's resource usage patterns regarding memory.
The gained information shall then be used to extrapolate memory usage to larger input data sizes.
We can then select a suitable cluster configuration, consisting of a node type and the scale-out, which fulfills memory requirements to avoid considerably costly memory bottlenecks.

\begin{figure}[htb]
    \centering
    \includegraphics[width=1\columnwidth]{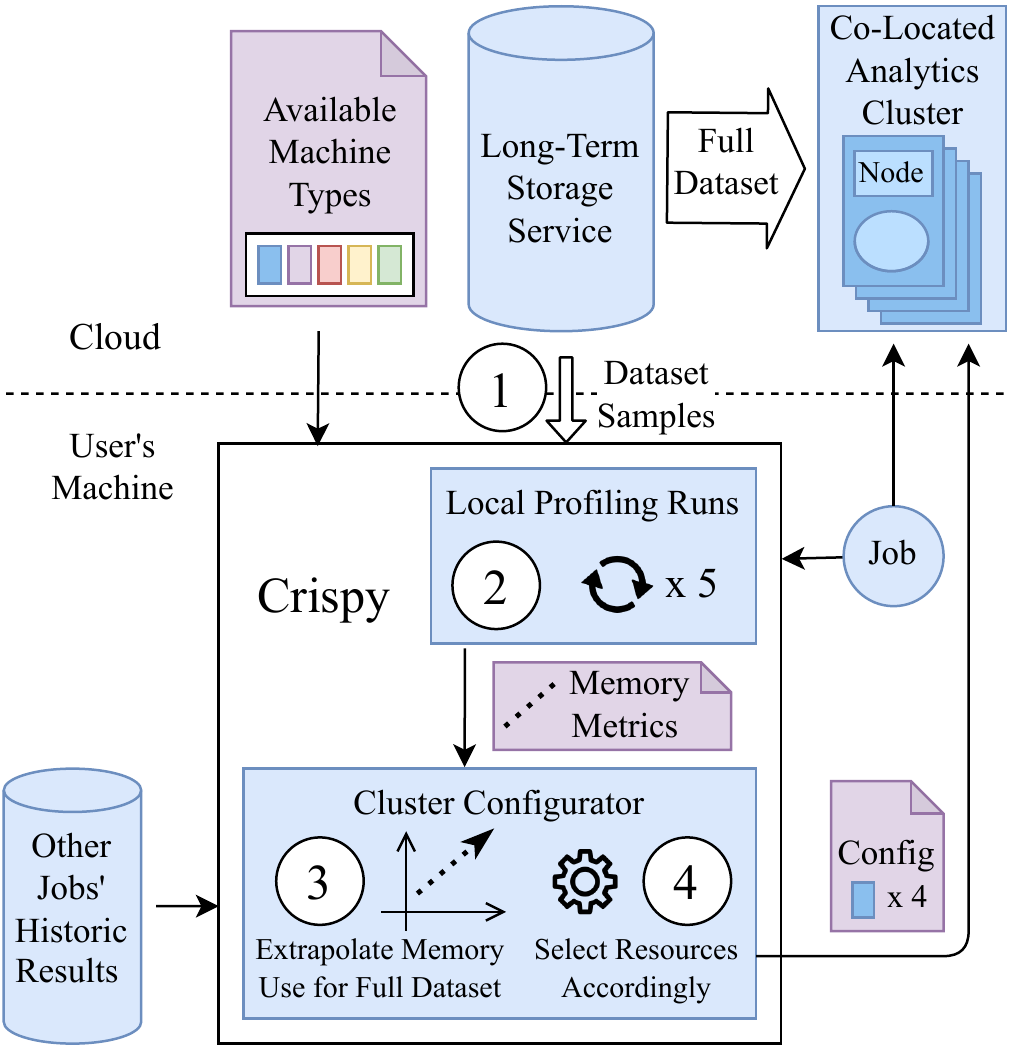}
    \caption{Overview of Crispy's components and their interaction.}\label{fig:crispy_overview}
\end{figure}

We propose a process for quickly and efficiently finding a good cluster resource selection, consisting of the following four steps that are visualized in Figure~\ref{fig:crispy_overview}:

\vspace{1.5mm}
\begin{tabularx}{\columnwidth}{lX}
    \circled{1} & \textbf{Dataset Sampling:} Load five small, differently sized samples of the dataset to a single node machine, e.g., the user's personal computer.\\\noalign{\vspace{1.5mm}}
    \circled{2} & \textbf{Profiling Runs:} Execute the job locally on each dataset sample and measure memory usage.\\\noalign{\vspace{1.5mm}}
    \circled{3} & \textbf{Memory Modeling:} Train a linear regression model to learn the relationship between data size and memory use.\\
                & \emph{If} the training $R^2$ score is high (> .99):\\
                & Assume that the modeling will be accurate and \\
                & estimate the total cluster memory requirement for \\
                & the execution on the full dataset,\\
                & \emph{Else}: Set the total cluster memory requirement to 0.\\\noalign{\vspace{1.5mm}}
    \circled{4} & \textbf{Configuration Selection:} Choose the historically most resource-efficient cluster configuration that fulfills this minimum total cluster memory requirement.\\\noalign{\vspace{1.5mm}}
\end{tabularx}

\subsection{Data Sampling and Job Profiling}

The job profiling is done on a single machine since our goal is not to model the entire execution, but merely how the amount of dataset bytes on storage translates to bytes in memory.
Therefore, this single machine does not need to be of the same type as the target cluster infrastructure, but can instead be a personal computer, thereby evading cluster resource occupancy.

The sizes of the data samples are chosen so that the execution time is just long enough for the startup process to have finished, and the actual job processing can take place for about half a minute to three minutes.
This provides enough time to measure the actual memory footprint of the dataset sample.
Initially, one percent of the original dataset can be chosen and then iteratively adjusted according to match those runtime targets, i.e., if the runtime is longer than three minutes, the profiling job can be canceled and restarted with a smaller portion of that sample.

Next, four more differently sized portions of this sample are used for additional profiling runs so that the sample sizes are equally spaced and reasonably far apart to then enable modeling and extrapolation.

There are several ways to continuously monitor memory usage for distributed dataflow jobs.
Ordered in descending specificity, the methods can be:
\vspace{1.5mm}
\begin{enumerate}
    \item Specific to the \emph{distributed dataflow framework}\\Example: The Spark history server\vspace{1.5mm}
    \item Specific to the \emph{programming language platform}\\Example: The Java Management Extensions (JMX)\vspace{1.5mm}
    \item Specific to the \emph{operating system}\\Example: \texttt{/proc/meminfo} in UNIX-like systems\vspace{1.5mm}
\end{enumerate}

For our proof of concept, which we evaluate in Section~\ref{sec:EVALUATION}, we use the approach of monitoring memory usage at the operating system level.
This is the most universal approach and is not specific to the distributed dataflow framework.
It further allows us to capture memory usage fluctuations for the job and also for the support systems that facilitate that job execution, e.g., the framework, distributed data storage system, and the resource managers.
This is also the measurement approach taken by the benchmarking tool \emph{HiBench}\footnote{\href{https://github.com/Intel-bigdata/HiBench}{github.com/Intel-bigdata/HiBench}, accessed March, 2022} by Intel, which we used in our evaluation.
To better assess the actual memory use of the job execution itself, the system-wide allocated memory before the start of execution is captured and accounted for.
Additionally, we tune the given garbage collector of the job's runtime to be more aggressive than its default settings which typically optimize for high throughput.
This gives us an idea of not simply how much of the memory is allocated by the job but how much memory is actually being used by the job at any given point.

A possible downside of capturing memory usage from this operating system interface is that the measurements can lose some accuracy, if there are unrelated programs with heavily fluctuating memory allocations on the same machine.
Therefore, co-locating such memory-intensive applications during profiling is not recommended when using this memory usage monitoring strategy.

\subsection{Memory Usage Modeling}

Our goal is to model the relation between input data size and used memory during job execution in order to estimate the required amount of memory for resource-efficient processing.
For this, we utilize the data points collected by the profiling runs on differently sized dataset samples.
We inspect the relation between data size and measured memory usage and examine whether the relation is linear or not.
This is achieved by training a linear regression model on the data.
If we observe a high accuracy on the training data itself, e.g., a score of $>.99$ when using the $R^2$ scoring metric, we assume the relation to be linear. Otherwise, we assume it is not.
We distinguish two cases:

\subsubsection{The relation is linear}

Those jobs are prone to memory bottlenecks since the memory use steadily grows with dataset size, and the cluster memory needs to accommodate this.
Due to the linear relation between dataset size and memory use, we can confidently extrapolate the job memory usage to the full dataset size.

\subsubsection{The relation is not linear}

Here, it will not be possible to model the memory requirement for an execution on the full dataset with our method.
Crispy then assumes the absence of memory bottlenecks and falls back to the most successful baseline approach, which will be presented in the evaluation in Section~\ref{sec:EVALUATION}.

Nonlinear relations could be measured for several different reasons.
For some jobs, the memory usage of a program is not related to the dataset size.
In this case, it is not necessary to actively avoid memory bottlenecks since there is no opportunity for them to occur.
For other jobs, there may be a linear relationship between the dataset size and its memory footprint, but the actual memory usage readings are obscured by rapidly generated objects with neither short nor endless lifespans.
The readings can also be affected by the possibly suboptimal active memory management through caching, as performed by the job programmer or the framework.  

\subsection{Selecting a Cluster Configuration}

To execute this job on the full dataset, we now want to select a suitable cluster resource configuration consisting of a node type and a scale-out.

Our baseline approach for selecting cluster resources to execute distributed dataflow jobs is what we call ``Best for All'' (BFA).
This approach works under our constraint that there are no existing runtime data for the job at hand.
To be able to compare different configurations, we need some notion of cost, for example, electric power consumption or monetary cost of renting the given cloud resources, which we use in our evaluation in Section~\ref{sec:EVALUATION}.
We then choose the configuration that has shown to be, on average, the most cost-efficient for all unrelated previous jobs.

In our approach, Crispy, we extend this baseline approach by considering only cluster configurations that fulfill the total cluster memory requirement, as determined by the memory usage modeling in the previous step.
We need to add a fixed amount of memory requirement per node to account for the overhead of the operating system and the dataflow framework, which is around two Gigabytes for Spark or Hadoop on an Ubuntu machine.
Here, it is also possible to add to the memory requirement as leeway to account for slight miscalculations or even phenomena like different Java Virtual Machine (JVM) implementations leading to slightly different footprints of objects in the JVM heap.

If we could not confidently model and extrapolate the memory use, the memory requirement is zero, in which case we are using exactly the BFA baseline approach.

%% file: sections/5_evaluation.tex
This section contains the evaluation of the Crispy cluster configurator in combination with its memory profiling aspect.
Specifically, the quality of the resource selections and the profiling time overhead are evaluated.

\subsection{Experimental Setup}\label{ssec:experimental_setup}

To conduct the local profiling runs, we used one co-author's work laptop, a 2020 T14 Thinkpad with 32~GB RAM and an AMD Ryzen 7 PRO 4750U CPU with up to 4.10~GHz.

\vspace{1.5mm}

\subsubsection{Prototype Implementation}

For the prototype implementation of the Crispy cluster configurator, we chose Python (version 3.10) for its code readability and its wealth of available libraries.
One such library, in particular, is \textit{Scikit-Learn} (v. 1.0.2)~\cite{scikit-learn}, which we benefited from when building the memory usage model.
Major supporting libraries we used were \textit{Numpy} (v. 1.22.0) and \textit{Pandas} (v. 1.3.5).
The local profiling experiments ran on Java 8, Spark 2.1.1, and Hadoop 2.7.3.

\vspace{1.5mm}

\subsubsection{Datasets}

\begin{figure*}[!b]
    \vspace{-4mm}
    \subfloat{%
      \includegraphics[width=.33\linewidth]{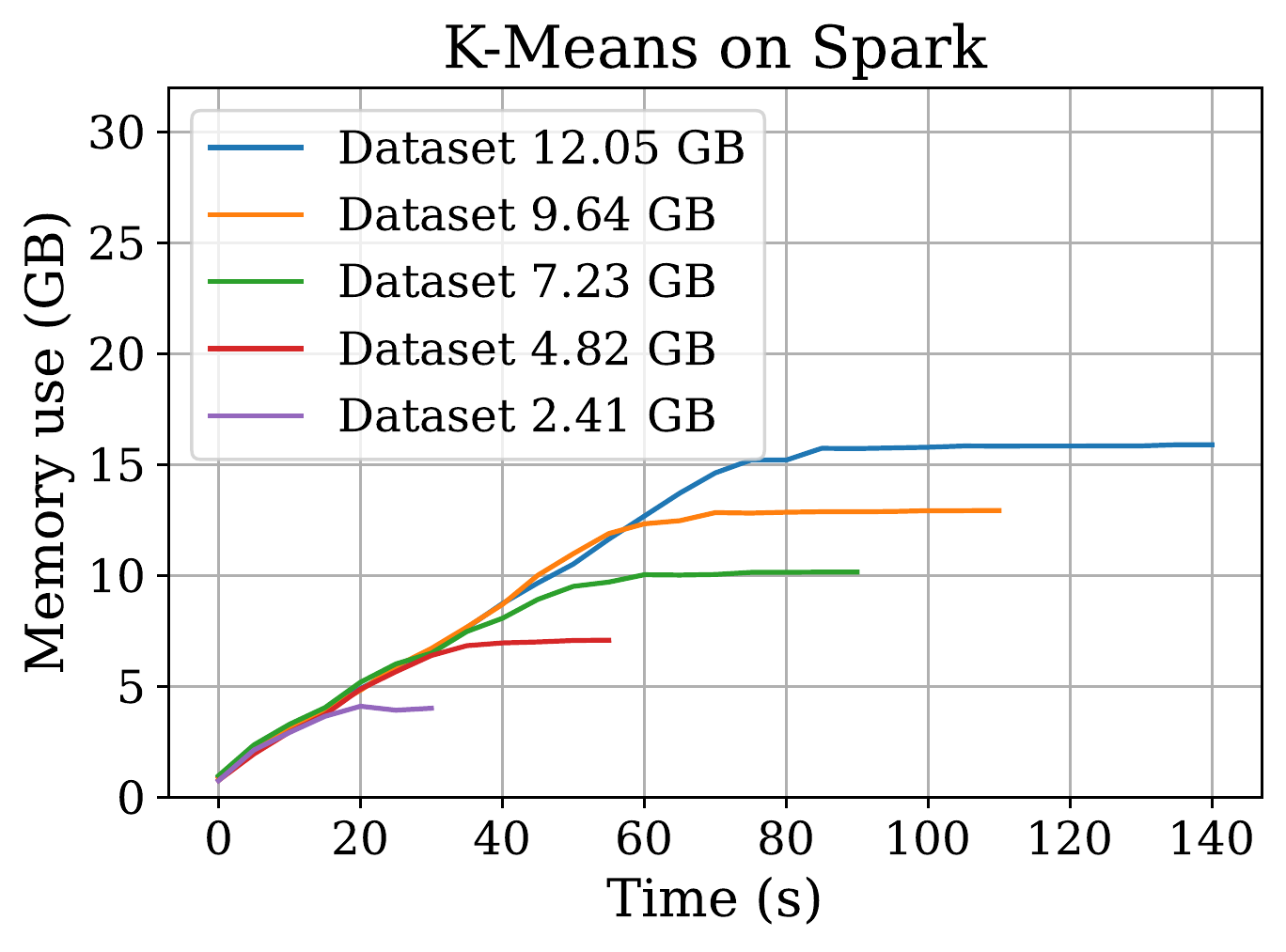}%
    }\hfill
    \subfloat{%
      \includegraphics[width=.33\linewidth]{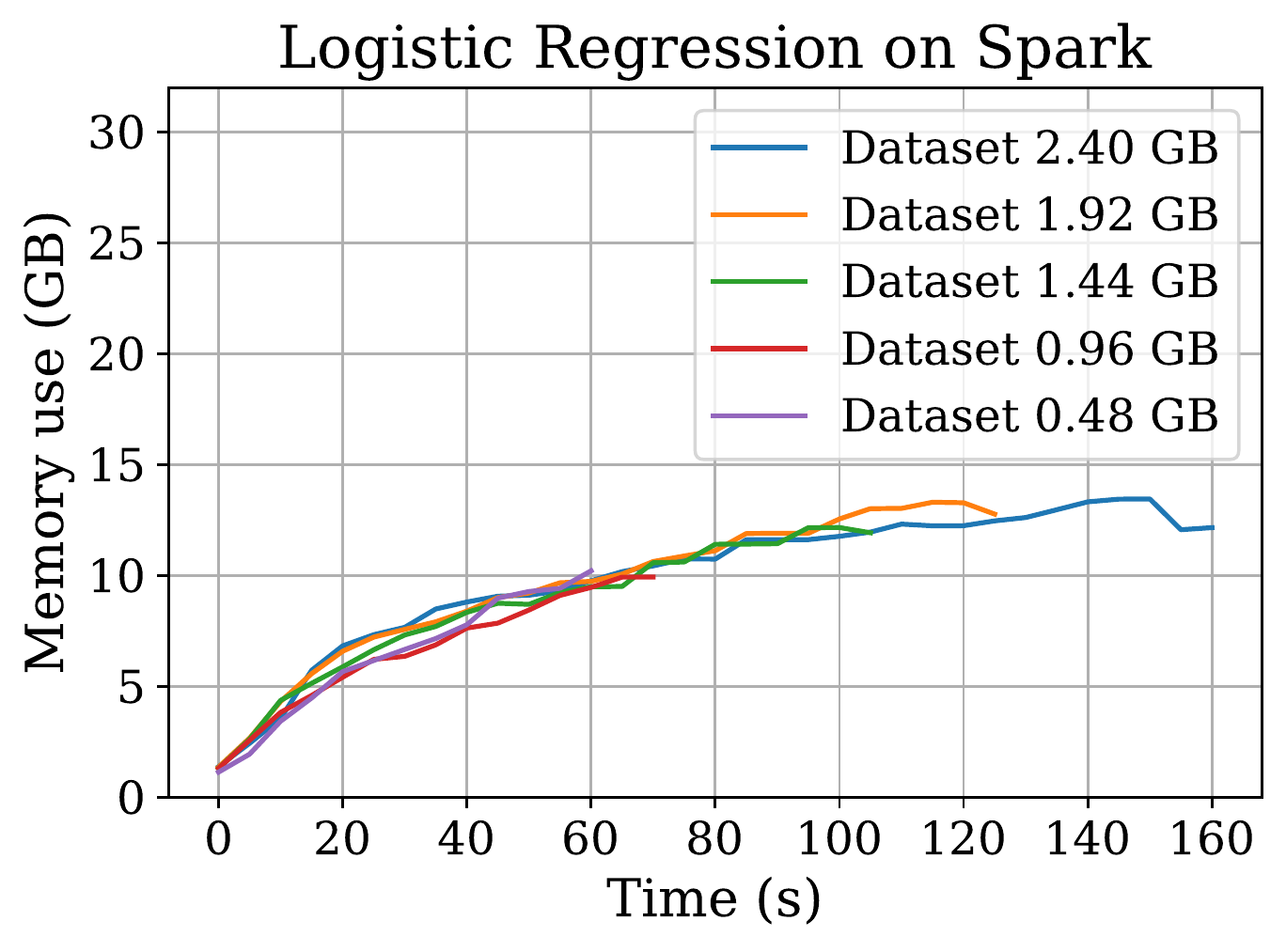}%
    }\hfill
    \subfloat{%
      \includegraphics[width=.34\linewidth]{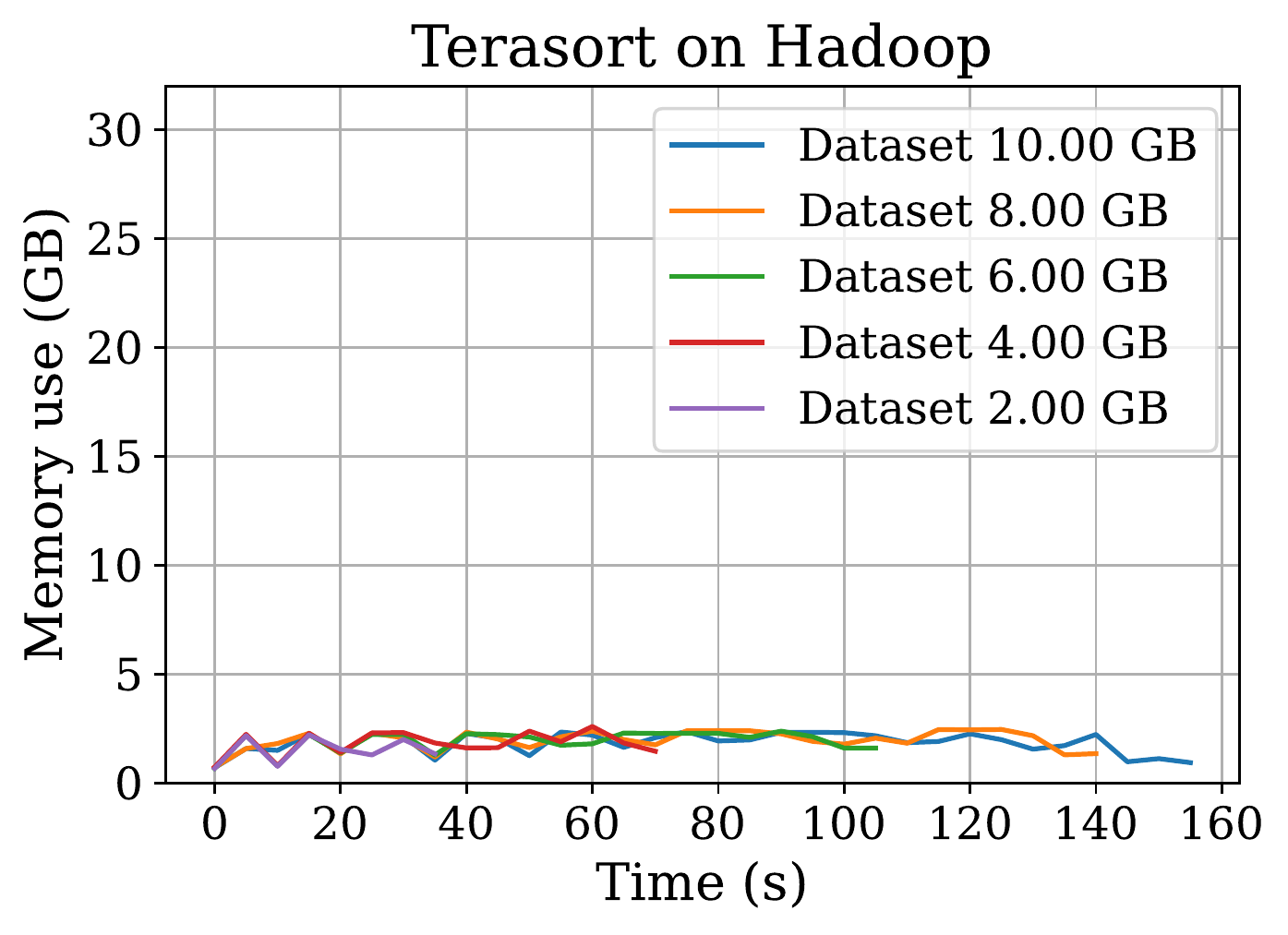}%
    }\hfill
    \caption{Memory use on a single-node machine measured over time for different jobs with aggressive garbage collection for newly created objects. Each job was executed with five linearly distributed dataset sample sizes.}\label{fig:memory_use}
\end{figure*}

For the evaluation of our methods, we chose an existing dataset\footnote{\href{https://github.com/oxhead/scout}{github.com/oxhead/scout}, accessed March 2022} that was used by Hsu et al.\ in \emph{Arrow}~\cite{hsu2018arrow}.
The dataset contains 1031 Spark and Hadoop unique executions on 69 different AWS cluster configurations, which were facilitated by the benchmarking tool \emph{HiBench} by Intel.
There are seven different underlying algorithms, and each job was executed with two different dataset sizes, with the smaller one being called ``huge'' and the larger being called ``bigdata''.

The cluster configurations have scale-outs between 4 and 48 machines and they have machine types of classes \emph{c}, \emph{m}, and \emph{r} in sizes \emph{large}, \emph{xlarge}, and \emph{2xlarge}.
Virtual machines of the \emph{c} type have less memory per core than those of the type \emph{r}, while machines of the \emph{m} type lie between those two.
Denominations like \emph{large}, \emph{xlarge} and \emph{2xlarge} refer to the amount of cores per machine.

\subsection{Memory Usage Measurement and Modeling}

Since the distributed data flow frameworks that we examine run on the JVM, they are subject to garbage collection.
To get a good idea of how much memory is actually being used, we need to make sure that unused objects are deleted from memory without substantial delay.
In our execution environment, we used the ``Parallel Garbage Collector'', which is the default garbage collector for Java 8.
For example, the parameter NewRatio denotes the relative size of the old generation of objects to the new generation of objects, with the default value being 2.33.
New objects that passed a few garbage collection audits are moved to the old generation, where they will only be evaluated again once this old generation is full.
Therefore, a smaller new generation leads to new objects being evaluated for garbage collection more quickly after their creation.
We passed this to the JVM as a runtime parameter.

\begin{figure}[bht]
    \begin{tabular}{@{}c@{}}
    \subfloat{%
      \includegraphics[width=.66\columnwidth]{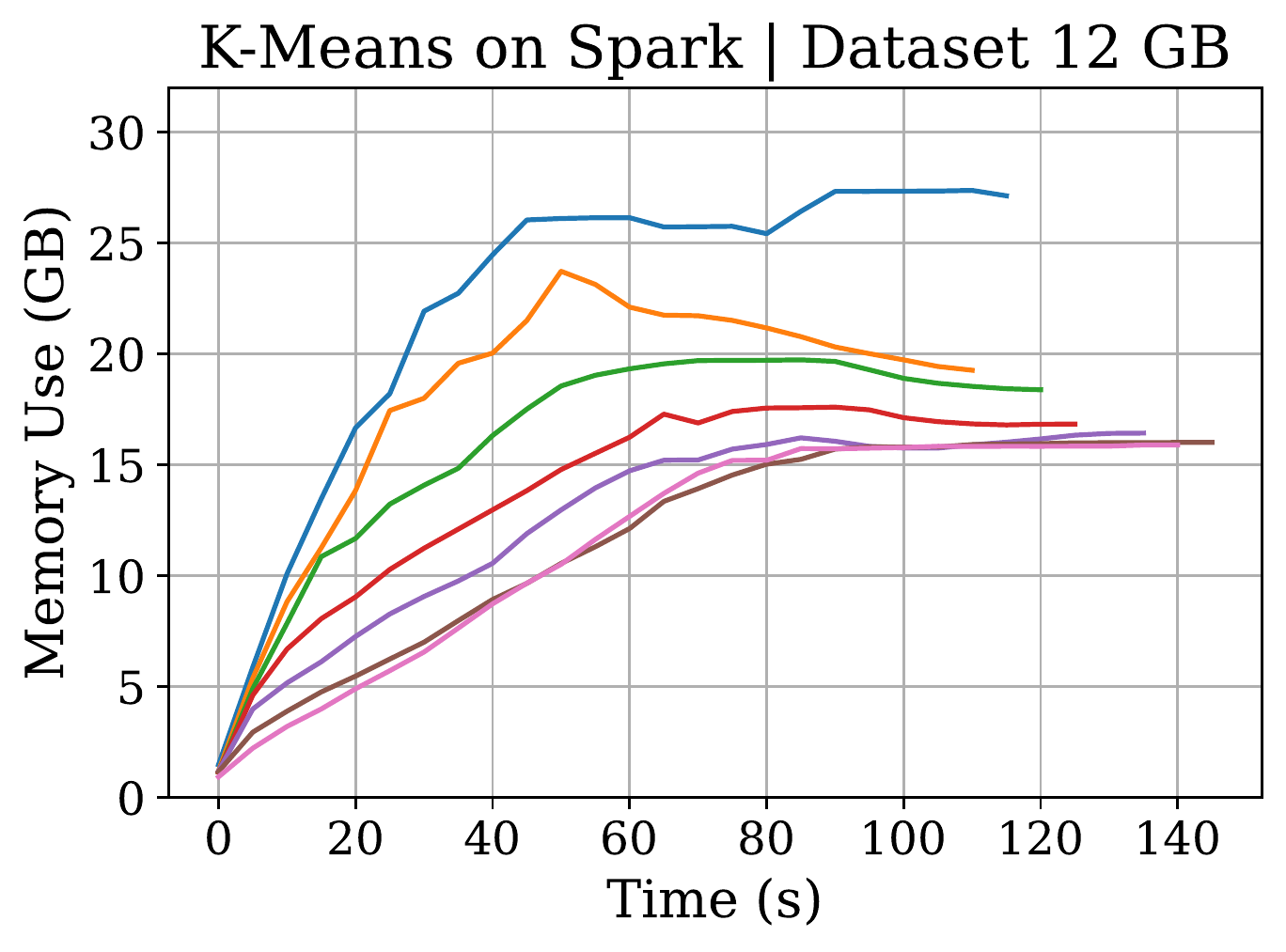}%
    }\hfill
    \end{tabular}
    \begin{tabular}{@{}c@{}}
    \subfloat{%
      \includegraphics[width=.33\columnwidth]{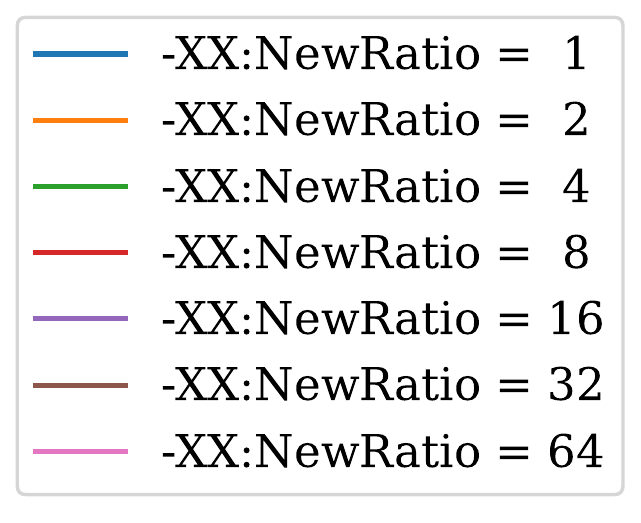}%
    }
    \end{tabular}
    \caption{Measured memory use over time for K-Means on a 12 GB dataset with different Java garbage collection settings. A higher NewRatio causes more aggressive garbage collection for newly created objects.}\label{fig:gc_settings}
\end{figure}

Figure~\ref{fig:gc_settings} shows how the aggressiveness of the garbage collection strategy influences the memory usage readings.
We see that a more aggressive garbage collection strategy allows us to get a clearer image of how much of the allocated memory is actually being used by a job.
At the same time, this leads to more time spent in garbage collection mode and, therefore, longer runtime, albeit to a degree that is still insignificant enough for our use case.

Figure~\ref{fig:memory_use} shows examples of the memory usage in different jobs.
With K-Means on Spark, we observe a linear relation between dataset size and peak memory use, while this is not the case for Logistic Regression on Spark and Terasort on Hadoop.
For K-Means, Crispy will thus extrapolate the memory use and, therefore, estimate the total cluster memory requirement to the full dataset.
For the other two jobs, Crispy does not attempt an extrapolation.

\subsection{Configuration Selection Quality}

To evaluate our selection strategy, we compare it to three baselines, all of which can be seen as approaches to selecting cluster resources for jobs without having access to runtime data from full historical executions.

We specifically investigate the monetary cost, since in public clouds like AWS, this is an adequate indicator of resource-efficiency.
For a job, the cost of each cluster configuration is normalized to the configuration that had the cheapest execution.
This means that the cheapest cluster configuration for a job always has a cost of 1.0, and another configuration that resulted in a three and a half times higher execution cost for this job would then have a cost of 3.5.
\vspace{6mm}\\
The baselines are:
\begin{enumerate}
    \item \emph{Random} -- The first baseline simulates the result of a randomly selected cluster configuration by taking the average cost.
    \item \emph{Medium} -- The second baseline approach is to always choose a medium-sized VM and a medium-sized scale-out for the cluster. In the case of our dataset, this is 12 machines of type \emph{m4.xlarge}.
    \item \emph{BFA} -- The third baseline approach stands for "Best For All" and selects the configuration that was, on average, the best for all previous jobs that were executed on the same distributed dataflow framework and that were different from the job at hand.
\end{enumerate}

Methods from related work do not qualify as baselines since their approaches assume job recurrence and thus being able to learn from full executions, which goes far beyond quick profiling on a single machine in terms of resource costs.
Crispy is capable of configuring resources for non-recurring, unique jobs without causing prohibitively large overhead.
This, along with further comparisons to related work, is elaborated in Section~\ref{sec:RELATED_WORK}.

\begin{table}[htb]
    \caption{Normalized Job Execution Cost by Configuration Selection Method for Crispy and Baselines}
    \vspace{-1mm}
    \begin{tabular}{l@{\hskip 1.3mm }l@{\hskip 1.98mm }l@{\hskip 1.95mm}cccc}
        && $\;\;\;\;\;\;\;$ & Random & Medium & BFA & Crispy  \\
        \hline\\[-1.7ex]
        \textbf{Naive Bayes}     & \textbf{Spark}   & bigdata &  1.2834  & 1.1731 &       1.0954  &      1.0954     \\
        \textbf{Naive Bayes}     & \textbf{Spark}   & huge    &  1.4083  & 1.3548 &       1.2039  &  \gn{1.0005}    \\\rowcolor{Gray}
        \textbf{K-Means}         & \textbf{Spark}   & bigdata &  3.4763  & 2.7873 &       3.9911  &  \gn{1.1570}    \\\rowcolor{Gray}
        \textbf{K-Means}         & \textbf{Spark}   & huge	  &  3.3398  & 3.1523 &       4.7778  &  \gn{1.1759}    \\
        \textbf{Lin. Regr.}      & \textbf{Spark}   & bigdata &  1.3531  & 1.2105 &       1.1334  &      1.1334     \\
        \textbf{Lin. Regr.}      & \textbf{Spark}   & huge	  &  3.1964  & 3.7181 &       3.1212  &      3.1212     \\\rowcolor{Gray}
        \textbf{Log. Regr.}      & \textbf{Spark}   & bigdata &  3.5475  & 2.5025 &       1.7318  &      1.7318     \\\rowcolor{Gray}
        \textbf{Log. Regr.}      & \textbf{Spark}   & huge	  &  5.2102  & 4.1047 &       2.4874  &      2.4874     \\
        \textbf{Page Rank}       & \textbf{Spark}   & bigdata &  1.4295  & 1.2261 &       1.3944  &  \gn{1.2469}    \\
        \textbf{Page Rank}       & \textbf{Spark}   & huge	  &  1.7973  & 1.3513 &       1.2040  &  \gn{1.0839}    \\\rowcolor{Gray}
                Join             & Spark            & bigdata &  1.8483  & 1.5673 &   \gn{1.0507} &  \gn{1.0507}    \\\rowcolor{Gray}
                Join             & Spark            & huge    &  2.5481  & -$^*$  &   \gn{1.0000} &  \gn{1.0000}    \\
                Page Rank        & Hadoop           & bigdata &  1.6641  & 1.4995 &   \gn{1.0000} &  \gn{1.0000}    \\
                Page Rank        & Hadoop           & huge    &  2.0419  & 1.8671 &   \gn{1.0000} &  \gn{1.0000}    \\\rowcolor{Gray}
                Terasort         & Hadoop           & bigdata &  1.6462  & 1.3631 &   \gn{1.1162} &  \gn{1.1162}    \\\rowcolor{Gray}
                Terasort         & Hadoop           & huge    &  1.7907  & 1.2695 &   \gn{1.0000} &  \gn{1.0000}    \\
        \hline\hline\\[-1.7ex]
        Mean                             &&&     2.3488  & 2.0098       & 1.7692 &      1.3375     \\
    \end{tabular}

    \vspace{3mm}
    \rule{\columnwidth}{.1mm}

    \vspace{1mm}

    $^*${\footnotesize Value missing from the original dataset}
    \label{tab:costs}
\end{table}

Table~\ref{tab:costs} shows the normalized cost of the configurations chosen by Crispy and each of the three baselines.
The jobs that are marked in bold are iterative Spark jobs, which by their nature, are prone to memory bottlenecks.

For Naive Bayes, K-Means, and Page Rank on Spark, Crispy successfully captured memory metrics in the profiling runs and modeled the memory usage for jobs with the full dataset.
Therefore, Crispy was able to mitigate severe memory bottlenecks and choose cluster configurations that are close to the optimum.
No improvement from the BFA baseline could be achieved for the Spark jobs Linear Regression and Logistic Regression, because Crispy could not extrapolate the memory requirement to the full dataset size with high enough confidence.
Here it fell back to the baseline approach, BFA\@.
In summary, Crispy was able to recognize and avoid the majority of memory bottlenecks for the jobs in our evaluation.

Overall, we see that Crispy's selections only incur costs that are, on average, around 34\% more costly than the perfect solution.
This constitutes an improvement of 56\% from the best-performing baseline, BFA\@.
For 75\% of the examined jobs, Crispy picks a configuration that is at most 20\% more costly than the optimum.

\subsection{Profiling Speed}

\begin{table}[htb]
    \centering
    \caption{Crispy's Profiling Time for all Jobs}
    \vspace{-1mm}
    \begin{tabular}{l@{\hskip 1.3mm }l@{\hskip 1.98mm }l@{\hskip 3.95mm}r}
        && $\;\;\;\;\;\;\;$ & {\hspace{3mm} Time (s)\hspace{-2mm}} \\
        \hline\\[-1.7ex]
            Naive Bayes & Spark  & bigdata & 373  \\
            Naive Bayes & Spark  & huge    & 369  \\\rowcolor{Gray}
            K-Means     & Spark  & bigdata & 470  \\\rowcolor{Gray}
            K-Means     & Spark  & huge	   & 470  \\
            Lin. Regr.  & Spark  & bigdata & 372  \\
            Lin. Regr.  & Spark  & huge	   & 198  \\\rowcolor{Gray}
            Log. Regr.  & Spark  & bigdata & 675  \\\rowcolor{Gray}
            Log. Regr.  & Spark  & huge	   & 562  \\
            Page Rank   & Spark  & bigdata & 1292 \\
            Page Rank   & Spark  & huge	   & 1292 \\\rowcolor{Gray}
            Join        & Spark  & bigdata & 136  \\\rowcolor{Gray}
            Join        & Spark  & huge    & 110  \\
            Page Rank   & Hadoop & bigdata & 812  \\
            Page Rank   & Hadoop & huge    & 812  \\\rowcolor{Gray}
            Terasort    & Hadoop & bigdata & 547  \\\rowcolor{Gray}
            Terasort    & Hadoop & huge    & 547  \\
        \hline\hline\\[-1.7ex]
            Mean        &        &         & 565  \\
    \end{tabular}
    \label{tab:profiling_times}
\end{table}

Table~\ref{tab:profiling_times} shows that profiling times were between two and 20 minutes, while the average was just below ten minutes.
The median is below eight minutes on the hardware as described in Subsection~\ref{ssec:experimental_setup}.

\subsection{Discussion}

Overall, the evaluation results indicate that Crispy provides a significant improvement from the best baseline and that the majority of memory bottlenecks were found and mitigated.
Crispy achieves this without knowledge from prior executions of the same job, incurring only small resource and time cost expenditures for executing the profiling runs.
Ten minutes of profiling on a personal computer is essentially free of cost, and the extent of this profiling effort is irrespective of the size of the full dataset.

In cases where the memory usage modeling and extrapolation do not work, e.g., the memory readings are not accurate enough, Crispy is able to recognize this and falls back to a baseline approach.
Therefore, Crispy has shown to be as good or better than the baseline approach for each of the 16 jobs in our evaluation.

%% file: sections/6_related_work.tex
Our system aims to be fairly agnostic to concrete data processing systems, which is why we devised a black-box job profiling and cluster configuration approach.
This section discusses related black-box approaches to cluster configuration.

\subsection{Approaches Based on Historical Performance Data}

Some approaches use runtime data to predict the job's scale-out and runtime behavior.
This data is gained either from dedicated profiling or previous full executions~\cite{ernest, hongzi2016resource, chen2021silhouette, rajan2016perforator, bell, scheinert2021enel, scheinert2021bellamy, scheinert2021potential, will2020towards, will2021c3o, will2021training}.
The models can then be used to predict the execution performance for different cluster configurations, and the most resource-efficient one will be chosen.
This can, for example, be the one with the lowest expected execution cost in a public cloud scenario under consideration of potential additional constraints.

\emph{Ernest}~\cite{ernest} trains a parametric model for the scale-out behavior of jobs on the results of sample runs on reduced input data, which works well for recurring programs that exhibit a rather intuitive scale-out behavior.
Initial configurations are tried out based on optimal experiment design.

With \emph{C3O}~\cite{will2021c3o}, the execution context is taken into consideration, and the effects of performance-influencing factors other than the cluster setup are modeled.
This enables learning from previous job executions even if they had vastly different parameter and dataset inputs than the current job.

Silhouette~\cite{chen2021silhouette} is a cloud configuration selection framework based on performance modeling with minimal training overhead.
Silhouette uses advanced statistical techniques and proposes a model transformer for quick transfer learning, and it effectively optimizes cloud configurations under constraints.

The disadvantage of all resource selection approaches based on performance models is that they either assume the availability of training data or they assume a recurring job where the overhead of profiling on the target infrastructure is offset by increased resource efficiency in future executions.
This disadvantage is aggravated by the need for models to be aware of the execution context.
Thus, the models can either only consider training data from equal contexts or they have to incorporate that context into their models, which requires even more training data to be available.

In contrast to these approaches, Crispy does not assume the existence of any performance metrics from previous job executions.
Metrics are purposefully and exclusively collected for the job in question.

\subsection{Approaches Based on Iterative Search}

Other approaches configure the cluster iteratively through profiling runs, attempting to find a better configuration at each iteration based on runtime information from prior iterations.
There are different strategies for reaching a stopping criterion, i.e., when it is expected that further exploration of the search space will not lead to significant enough exploitation potential to justify the additional search overhead~\cite{cherrypick, hsu2018micky, hsu2018arrow, fekry2020accelerating, he2019statistics}.

\emph{CherryPick}~\cite{cherrypick} tries to directly predict the optimal cluster configuration that obeys given runtime targets by utilizing Bayesian optimization.
The search stops once it has found the optimal configuration with reasonable confidence.

In \emph{Arrow}~\cite{hsu2018arrow}, the authors use low-level metrics such as memory utilization to enhance previously established Bayesian optimization methods.
These metrics are, for instance, CPU utilization, working memory size, and I/O wait time.
Utilizing them allows for finding optimal or near-optimal configurations with fewer iterations.

\emph{SimTune}~\cite{fekry2020accelerating} combines workload characterization and Multitask Bayesian optimization to speed up its incremental search for near-optimal configurations.
Its setup allows the tuning to be done online and proves to significantly reduce the exploration costs for finding close-to-optimal configurations.

These highlighted methods assume that the search cost is offset by resource efficiency in subsequent executions of a recurring job.
However, it is questionable how well the configurations found by these methods can be applied for subsequent jobs that have slightly different inputs.
For instance, previously found configurations might lead to, e.g., memory bottlenecks for subsequent jobs.

\emph{PT4Cloud}~\cite{he2019statistics} iteratively builds performance models for a job on different cloud configurations with data from profiling runs.
Users define an accuracy requirement, allowing for a tradeoff between model accuracy and search time.
In accordance with this accuracy requirement, the stopping condition is based on the statistical stability of the performance model.
PT4Cloud adds training data until the performance models are considered statistically stable to avoid unnecessary exploration.

Compared to these approaches, our solution is based on profiling only the job itself to establish memory usage patterns.
Therefore, Crispy does not need to run the profiling runs on the full target cluster but instead can do so on a single machine, i.e., just a single node in the target cluster or on the developer's laptop.
This, in turn, saves cluster resources and enables finding an optimized configuration for every individual job execution.
Therefore, Crispy can optimize resource configurations of unique, one-off jobs.

\subsection{Cross-Infrastructure Performance Prediction}

Futher approaches attempt to predict data processing job performance based on performance metrics previously measured on diverse infrastructures~\cite{yang2005cross, sadjadi2008modeling, baughman2018profiling, yadwadkar2017selecting, scheinert2021bellamy}.

\emph{Baughman et al.}~\cite{baughman2018profiling} use a two-fold approach.
They can make use of previously gathered runtime data of the same job, which may have been executed on different infrastructures, and use transfer learning to train performance models for the target infrastructure.
If such data from previous executions is not available, they resort to profiling runs that record metrics like CPU and Memory utilization.

\emph{PARIS}~\cite{yadwadkar2017selecting} attempts to model the performance cost trade-off for a given set of VM types to let the user decide based on preference.
For this, PARIS profiles a set of cloud VM types with a reference job that has diverse resource requirements.
To choose an optimized cloud configuration for an actual production job, this job is executed on a set of reference VMs.
PARIS then attempts to infer the job performance on all other VM types, which saves effort compared to actual executions.

\emph{Bellamy}~\cite{scheinert2021bellamy} captures the context of a job execution, including node properties, scale-outs, and dataset sizes.
Bellamy then realizes a two-step modeling approach.
First, a general model is trained on all the available data for a specific scalable analytics algorithm, hereby incorporating data from different contexts.
Second, the general model is optimized for the specific situation, based on the available data for the concrete context, which can even be on a new cluster.

In summary, the presented approaches for cross-infrastructure performance prediction differ from Crispy in two main ways.
On the one hand, they assume that the job is recurring in some form, which allows them to learn from full previous executions, but makes them also not applicable to unique, one-off jobs, which Crispy is primarily concerned about.
On the other hand, they disregard the occurrence and relevance of memory bottlenecks.
The presence of a memory bottleneck for a job results in cluster configurations behaving much differently depending on the input dataset size, which is why Crispy is centered around this problem.
Hence, these approaches lose their efficacy once either the assumption of job recurrence or the assumption of input dataset size constancy does not hold true.

%% file: sections/7_conclusion.tex
This paper presented Crispy, a system that allows users to profile a distributed dataflow job on a single machine and use the gained knowledge about the job's memory requirements to select a suitable cluster configuration that avoids costly memory bottlenecks.

In our experimental evaluation, we see Crispy circumventing the majority of memory bottlenecks by successfully modeling memory use in relation to input dataset size.
For the other cases, where our current methods of measuring and modeling memory consumption fall short, we show that Crispy reliably falls back to the baseline approach.
In total, there is a reduction of job execution costs by 56\% from our baseline.
Crispy spent an average of less than ten minutes for job profiling runs on a consumer-grade laptop.

Compared to state-of-the-art methods, our approach can work without the assumption of job recurrence and is thus suitable for optimizing resource selections for unique, one-off data processing jobs.

Future work may include using knowledge about memory bottlenecks to speed up iterative approaches to configure cloud resources for recurring data processing jobs.